\documentclass[12pt]{article}
\usepackage{amsmath}
\usepackage{graphicx}
\usepackage{natbib}
\usepackage{url} 
\usepackage{caption}

\newcommand{\blind}{0}

\begin{document}

\def\spacingset#1{\renewcommand{\baselinestretch}%
{#1}\small\normalsize} \spacingset{1}


\if0\blind
{
  \title{\bf Adaptive Physics-Guided Neural Network}
  \author{David Shulman\hspace{.2cm}\\
    and \\
    Itai Dattner \\
    Department of Statistics, University of Haifa, Haifa, Israel}
  \maketitle
} \fi

\if1\blind
{
  \bigskip
  \bigskip
  \bigskip
  \begin{center}
    {\LARGE\bf Adaptive Physics-Guided Neural Network}
\end{center}
  \medskip
} \fi

\bigskip
\begin{abstract}

This paper introduces an adaptive physics-guided neural network (APGNN) framework for predicting quality attributes from image data by integrating physical laws into deep learning models. The APGNN adaptively balances data-driven and physics-informed predictions, enhancing model accuracy and robustness across different environments. Our approach is evaluated on both synthetic and real-world datasets, with comparisons to conventional data-driven models such as ResNet. For the synthetic data, 2D domains were generated using three distinct governing equations: the diffusion equation, the advection-diffusion equation, and the Poisson equation. Non-linear transformations were applied to these domains to emulate complex physical processes in image form.

In real-world experiments, the APGNN consistently demonstrated superior performance in the diverse thermal image dataset. On the cucumber dataset, characterized by low material diversity and controlled conditions, APGNN and PGNN showed similar performance, both outperforming the data-driven ResNet. However, in the more complex thermal dataset, particularly for outdoor materials with higher environmental variability, APGNN outperformed both PGNN and ResNet by dynamically adjusting its reliance on physics-based versus data-driven insights. This adaptability allowed APGNN to maintain robust performance across structured, low-variability settings and more heterogeneous scenarios. These findings underscore the potential of adaptive physics-guided learning to integrate physical constraints effectively, even in challenging real-world contexts with diverse environmental conditions.

\end{abstract}

\noindent%
\vfill

\newpage
\spacingset{1.75} 

\section{Introduction}

Inverse problems are prevalent across various scientific and engineering disciplines, including geophysics, medical imaging, and material science. The aim is to infer unknown parameters or physical properties from observed data, often governed by underlying physical models. Unlike well-posed forward problems, inverse problems can be computationally challenging due to their inherent sensitivity to data variations and the complexity involved in estimating the underlying parameters. While traditional approaches such as regularization and statistical modeling provide solutions \cite{kaipio2007statistical}, these methods can become inefficient when dealing with high-dimensional and complex systems \cite{lieberman2010parameter}.

The advent of deep learning has introduced powerful new tools for addressing these challenges in inverse problems, \cite{de2022deep}. Neural networks, particularly deep architectures, are capable of modeling complex and nonlinear relationships between inputs and outputs, which makes them well-suited for high-dimensional problems. For instance, \cite{10.1214/18-AOS1747} demonstrated that deep neural networks could effectively circumvent the curse of dimensionality in nonparametric regression, provided suitable network architectures are employed. This flexibility has led to the application of deep learning models in various domains, including classification tasks involving multidimensional functional data \cite{10.1214/24-EJS2229} and image-based predictions \cite{10.1214/22-AOAS1637}.

Despite the effectiveness of deep learning models in many applications, a significant challenge remains in ensuring the stability and robustness of these models when applied to inverse problems. Traditional deep learning models can be sensitive to noise and small perturbations in the input data, which may lead to incorrect predictions and reduced reliability \cite{10.1214/22-AOAS1637}. To address these concerns, researchers have begun integrating domain-specific knowledge, particularly through the development of Physics-Informed Neural Networks (PINNs). PINNs incorporate physical laws directly into the network’s architecture or loss function, ensuring that the solutions generated by the model adhere to known physical constraints. This hybrid approach enhances the interpretability and reliability of deep learning models when applied to physical systems, as demonstrated by \cite{del2023deep}.

In parallel, the use of adaptive neural networks has gained traction as a means of improving the flexibility and efficiency of deep learning models. Adaptive neural networks dynamically adjust their structure based on the complexity of the data, enabling them to efficiently capture the underlying patterns without overfitting or becoming computationally prohibitive. For instance, \cite{Adaptive_Neural} introduced a neural network framework that optimizes the network structure according to the underlying data’s features, leading to improved estimation accuracy and performance. This approach builds on earlier work by \cite{10.1214/aos/1176347963}, which proposed multivariate adaptive regression splines as a flexible method for high-dimensional data modeling. Such adaptive techniques are particularly valuable when integrating neural networks with domain-specific knowledge, as they allow for efficient learning while maintaining interpretability.

Furthermore, combining adaptive methods with physics-guided approaches has the potential to optimize model performance in scenarios where domain knowledge can significantly inform the modeling process. Adaptive algorithms have been shown to effectively reduce model complexity, allowing for targeted and efficient solutions, as evidenced by \cite{10.1155/2013/452653}. This makes them well-suited for applications where the structure of the data or the complexity of the physical system can vary across domains, necessitating a flexible yet accurate modeling framework.

In our previous work \cite{shulman2024physics}, we introduced a Physics-Guided Neural Network (PGNN) framework for inverse regression, specifically designed for crop quality assessment. In that study, we integrated physical laws such as Fick’s second law of diffusion into the neural network architecture to predict quality metrics for agricultural products. This approach significantly improved the accuracy of predictions compared to conventional methods by incorporating the underlying physics of the problem into the learning process.

Our previous study applied the PGNN model to a single dataset and demonstrated that solving the inverse problem first by inferring physical properties before proceeding to the forward prediction task led to better overall performance. This physics-guided approach allowed us to make more accurate predictions based on inferred physical properties, providing a scalable and practical solution to agricultural quality assessment.

Building on the foundation laid in our previous work, the current study extends the PGNN framework by introducing an Adaptive Physics-Guided Neural Network (APGNN) approach that incorporates weight prediction. The goal of this work is to solve the inverse problem first, using unsupervised learning to infer physical properties from observed image data. After solving the inverse problem, a forward model is used to predict observable outcomes, such as quality metrics or classification labels. This approach is further enhanced by the use of an adaptive structure within the neural network, allowing the model to dynamically adjust its complexity based on the problem at hand.

One of the key innovations in this study is the integration of a combined weights solution, where predictions from the PGNN are fused with those from a traditional data-driven neural network (ResNet). This hybrid approach combines the strengths of physics-based learning with purely data-driven methods, ensuring that the model benefits from both the underlying physical principles and the flexibility of deep learning.

In contrast to our previous work, which was limited to a single type of dataset, the current study applies this adaptive physics-guided neural network framework to two distinct real-world datasets, each representing different physical phenomena:
\begin{enumerate}
    \item Cucumber Quality Assessment using RGB images, where the task is to predict continuous quality metrics based on physical properties.
    \item Thermal Camera Material Classification, which involves classifying indoor and outdoor materials based on thermal images.
\end{enumerate}

In addition to these real-world datasets, we conduct a simulation study involving various governing physical equations (e.g., diffusion, advection-diffusion, Poisson equations) and non-linear mapping functions that transform the physical properties into image representations. This allows us to evaluate the performance of our approach across a range of physical systems and data complexities, providing valuable insights into the effectiveness and generalizability of the Adaptive Physics-Guided Neural Network approach.

Our study offers a comparison between the adaptive physics-guided neural network (APGNN), physics-guided neural network (PGNN) and a conventional data-driven ResNet model, assessing their performance across different training sizes, governing equations, and real-world scenarios. Our aim is to demonstrate that adaptive, combined prediction within the APGNN framework improves prediction accuracy across a wide range of scenarios. By dynamically balancing physics-guided and data-driven insights, APGNN consistently enhances performance, whether in controlled environments with limited material diversity or in more complex, noisy settings. This adaptive approach allows APGNN to achieve robust, high-quality predictions, overcoming the limitations often faced by models that rely solely on physical constraints or purely data-driven methods.

The contributions of this paper are threefold:
\begin{enumerate}
    \item We develop a adaptive physics-guided neural network framework that integrates physical laws into the learning process, enhancing prediction accuracy for image-based tasks.
    \item We provide a thorough evaluation of the APGNN approach on both synthetic and real-world datasets, comparing its performance with a conventional ResNet model across various scenarios.
    \item We identify the conditions under which adaptive physics-guided learning offers advantages and discuss its limitations, offering insights into the potential for combining physics-based constraints with data-driven models.
\end{enumerate}

The remainder of this paper is organized as follows: Section 2 outlines the methodology, explaining the governing equations and the APGNN framework. Section 3 details the implementation using two real-world datasets: vegetable quality assessment through RGB images and material classification with thermal images. Section 4 presents the simulation study along with a discussion of the results. Section 5 focuses on the real-world experiments, including an analysis of the outcomes. Finally, Section 6 concludes the paper, offering insights and suggestions for future research.

\section{Methodology}
\subsection{Problem Definition}

To maintain generality and clarity, we initially present our methodology without specifying the explicit dimensions of each variable. These details will be clarified in subsequent sections where the theoretical framework is applied to specific cases.

Consider a latent variable \( x \) representing an intrinsic characteristic or state of a system, which drives the observable outcomes of interest. The primary goal is to model the relationship between \( x \) and an associated observable quantity \( y \). This relationship can be formalized as:
\[
y = f(x),
\]
where \( f \) is an unknown functional mapping from a space of square-integrable functions to real numbers. In various applications, \( y \) might represent a measurable attribute derived from \( x \), such as sensor readings, experimental measurements, or any response variable of interest.

For a series of observations \( Y_i \), we propose the following model:
\[
Y_i = f(x_i) + \eta_i, \quad i = 1, \ldots, n,
\]
where \( Y_i \) denotes the observed measurement for the \(i\)-th sample. In this model, \( f(x_i) \) represents the true underlying relationship, while \( \eta_i \) captures the noise inherent in the measurement process. The error terms \( \eta_i \) are assumed to be independently and identically distributed (i.i.d.) as Gaussian random variables with mean zero and constant variance.

Moreover, let \( z \) denote another observable variable capturing supplementary information about the system, such as visual features, proxy measurements, or other forms of auxiliary data. This variable can be modeled as a function of the latent state \( x \):
\[
z = g(x),
\]
where \( g \) is a transformation mapping the intrinsic state \( x \) into the observed data \( z \).

In our framework, \( x \) is the underlying driving force that governs the observable phenomena, serving as the fundamental entity from which both \( y \) and \( z \) are derived. The data acquisition process for the observable \( z \) can be modeled as:
\[
Z_i = g(x_i) + \epsilon_i, \quad i = 1, \ldots, n,
\]
where \( \epsilon_i \) are i.i.d. Gaussian noise variables with mean zero and constant variance \( \sigma^2 \).

The primary objective is to leverage the paired data \( (Z_i, Y_i) \) for \( i = 1, \ldots, n \), in conjunction with a dynamic model for \( x \), to accurately learn the function \( f \). This approach integrates theoretical knowledge of the latent system dynamics with empirical observations, thus enhancing our predictive modeling capabilities.

The rationale for learning \( f \) stems from the need to understand and model the fundamental relationships within the system, particularly in scenarios where only observations \( Z \) are available to predict \( Y \). This endeavor goes beyond simple statistical modeling; it seeks to uncover the deeper scientific relationships governing the observed data.


\subsection{Integration of Physical Models in Latent State Analysis}

Our study examines an intrinsic characteristic \( x \) that evolves according to certain physical principles over time or space. These principles govern the behavior and evolution of \( x \), ensuring that it serves as a reliable indicator of the system's state or quality.

\paragraph{Modeling Framework:} To capture the dynamics of \( x \), we employ a differential equation framework, which encapsulates the influence of underlying physical laws. This is expressed as:
\begin{equation}\label{diff_operator}
\frac{\partial x}{\partial t} = \mathcal{P}(x, \nabla x, \nabla^2 x, \ldots),
\end{equation}
where \( \mathcal{P} \) is a differential operator representing the combined effect of various physical phenomena influencing \( x \). This operator may include elements of diffusion, advection, reaction kinetics, or other relevant dynamics, depending on the specific nature of the system under study.

The choice of \( \mathcal{P} \) is guided by theoretical insights and empirical data, and it serves as a general representation of the physical processes that affect \( x \). This formulation allows us to incorporate domain knowledge into the modeling of \( x \), providing a physically-informed foundation for the subsequent analysis.

\subsection{Conceptual Framework for Physically-Informed Mapping}

At the core of our methodology lies the relationship between the observable data \( z \) and the intrinsic state \( x \). We introduce a conceptual framework that defines a mapping between these entities, which is informed by the underlying physical laws governing the system.

\textbf{Methodological Approach - Mapping from \( z \) to \( x \)}: Our approach centers around the function \( g^{-1} \), representing an inverse mapping that reconstructs the latent state \( x \) from the observable data \( z \). This mapping is informed by the same physical principles that govern the transformation \( g \). By employing \( g^{-1} \), we aim to convert the observed features in \( z \) into quantifiable and meaningful representations of the intrinsic state \( x \).

This framework supports the development of models and algorithms that leverage auxiliary data for estimating and predicting the latent variable \( x \). It emphasizes the insight that the observable characteristics in \( z \) contain valuable information about the system's intrinsic state, which can be effectively harnessed through the mapping \( g^{-1} \). This enables a nuanced and accurate reconstruction of \( x \) from indirect observations, thus enhancing our understanding and prediction of the system's behavior.

\subsubsection*{Formulation of the Optimization Problem}

In our physics-informed framework, we aim to construct a mapping function \( \hat{g}^{-1} \) that effectively transforms the observed data \( Z_i \) into an estimated latent state \( \hat{x}_i \), guided by the underlying physical laws governing the system. A distinctive feature of this approach is its ability to operate without relying on labeled data; instead, the model learns directly from the input data \( Z_i \), utilizing the governing physical principles to inform the learning process. Our objective is for the estimated state \( \hat{x} \) to approximate the true latent state \( x \) as closely as possible. We anticipate achieving this with a sufficiently representative and diverse dataset, such that the discrepancy \( d(\hat{x}, x) \) is below a predetermined threshold \( \delta \).

The optimization problem can be formulated as:
\[
    \min_{\hat{g}^{-1}} \sum_{i=1}^{n} \mathcal{L}_{physical}(\hat{g}^{-1}(Z_i)),
\]
where \( \mathcal{L}_{physical} \) represents a physics-informed loss function. This loss function evaluates the degree to which \( \hat{g}^{-1}(Z_i) \) conforms to the physical model, based on criteria derived from the system's governing physical laws. The resulting physics-constrained model is capable of learning effectively in the absence of labeled data, guided solely by the intrinsic structure and principles of the system.

\subsection{Predictive Modeling from Observed Data to Target Measurements}

Our methodology involves predicting target measurements \( Y_i \) from the observed data \( Z_i \). We propose three principal strategies for this prediction task:

\paragraph{Method 1: Direct Prediction}
The first strategy entails developing a predictive model, denoted \( \hat{h} \), that establishes a direct mapping from the observed data \( Z_i \) to the target measurements \( Y_i \).

\textbf{Model \( \hat{h} \) - Direct Prediction:} This model is trained to take \( Z_i \) as input and output an estimated value for \( Y_i \). The construction of this model relies on a training dataset containing paired instances of \( Z_i \) and \( Y_i \).

The direct prediction method is efficient and effective when a strong correlation exists between the observed data and the target measurements. However, it may not fully exploit the deeper physical principles governing the system.

\paragraph{Method 2: Inverse Prediction via Estimated Latent State}
The second approach utilizes the physics-informed mapping function \( \hat{g}^{-1} \) to estimate the latent state \( \hat{x} \) from \( Z_i \). Subsequently, a separate model, \( \hat{f} \), is employed to predict \( Y_i \) based on the estimated state \( \hat{x} \).

\textbf{Model \( \hat{f} \) - Inverse Prediction:} The model \( \hat{f} \) is designed to use \( \hat{x} \), estimated from \( Z_i \), as input to predict \( Y_i \). The training of this model is performed on a dataset consisting of paired instances of estimated values \( \hat{x}_i \) and actual measurements \( Y_i \).

\paragraph{Effectiveness of the Inverse Prediction Approach}
The success of the inverse prediction method relies on two key factors:
\begin{itemize}
    \item The accuracy of the mapping \( \hat{g}^{-1}: Z \rightarrow \hat{x} \), measured by the discrepancy \( d(\hat{x}, x) \).
    \item The precision of the model \( \hat{f}: \hat{x} \rightarrow Y \) in mapping \( \hat{x} \) to \( Y_i \), evaluated using the metric \( d(\hat{Y}_i, Y_i) \).
\end{itemize}

We hypothesize that this two-stage inverse prediction method may yield more accurate results than the direct approach \( \hat{h}: Z \rightarrow Y \), particularly when the cumulative error (\( d(\hat{x}, x) \) and \( d(\hat{Y}, Y) \)) is less than the error of the direct model. This approach is especially advantageous in situations where \( \hat{x} \) captures complex, latent characteristics that are not directly observable in \( Z \), enabling a more nuanced and physically-informed prediction.

\paragraph{Method 3: Adaptive Prediction Approach}

The third approach combines both the direct and inverse prediction strategies, allowing us to leverage the strengths of each method. The goal is to create a adaptive prediction model that balances the direct model's efficiency with the inverse model's physically-informed accuracy.

\textbf{Adaptive Prediction Model:} The final predicted value for \( Y_i \), denoted \( \hat{y}_i \), is obtained as a weighted combination of the predictions from the direct model \( \hat{h}(Z_i) \) and the inverse model \( \hat{f}(\hat{g}^{-1}(Z_i)) \):
\[
\hat{y}_i = \omega_i \hat{h}(Z_i) + (1 -\omega_i) \hat{f}(\hat{g}^{-1}(Z_i)),
\]
where \( \omega_i \in [0, 1] \) is a weight that determines the relative contribution of the two models.
We employed dynamic weighting, where \( \omega_i \) is adapted as a function of the input features \( Z_i \).

\paragraph{Advantages of the Adaptive Prediction Approach}
The adaptive prediction model offers flexibility, combining the predictive efficiency of the direct method with the physically-informed accuracy of the inverse method. When the direct model \( \hat{h}(Z_i) \) is highly effective, a larger \( \omega \) can be applied. Conversely, in scenarios where the inverse method provides more detailed insights, a smaller \( \omega \) enhances the model’s accuracy.

\paragraph{Comparative Performance Evaluation}
To evaluate the performance of the adaptive prediction model, we employ a validation strategy that compares the model’s predictions using standard metrics such as RMSE or Mean Absolute Error (MAE). The combined model’s performance \( \hat{y}_i \) is benchmarked against the individual predictions from both the direct \( \hat{h}(Z_i) \) and inverse \( \hat{f}(\hat{g}^{-1}(Z_i)) \) methods. This comparative analysis, conducted across multiple validation sets, provides robust evidence for the efficacy of the adaptive prediction strategy.

We hypothesize that this combined approach will yield superior predictive accuracy compared to standalone models, particularly in contexts where neither model consistently outperforms the other across the entire dataset. The flexibility of this adaptive approach enables it to adapt to varying data conditions, leveraging both direct and physics-informed insights to achieve optimal prediction accuracy.
\section{Implementation}
In this section, we present the application of our physics-guided framework using two real-world datasets. The first dataset involves assessing vegetable quality through RGB images, while the second dataset focuses on the classification of indoor and outdoor materials using thermal images. Both cases demonstrate how integrating physical laws into machine learning models can enhance prediction accuracy and interpretability.

\subsection{Methodological Adaptation for Vegetable Images}
\subsubsection{Overview}
The first case study involves using RGB images to assess the quality of vegetables by relating image data to laboratory measurements and understanding changes in moisture content. The freshness and overall quality of vegetables, such as cucumbers, are closely tied to their moisture content. As they age or are stored under suboptimal conditions, moisture loss becomes a significant factor, leading to a decline in quality attributes such as texture, weight, and appearance \cite{cantor2020equations, omolola2017quality, yahia2018postharvest, onwude2016modeling}. Accurate modeling of moisture dynamics is therefore crucial for quality assessment.

To model this accurately, we employ Fick’s second law of diffusion, which provides a physical basis for understanding and simulating moisture movement within the vegetables. By integrating Fick’s law into our algorithm, we can effectively relate the spatial distribution of moisture content to the overall quality attributes captured in the RGB images. Neural networks (NNs) are utilized to capture these non-linear relationships, making them particularly suitable for this application, as they can map complex dependencies between the image features and quality metrics, all while adhering to the underlying physical principles.
The dataset utilized in this work was obtained from materials provided in the original manuscript \cite{shulman2024physics}.
\subsubsection{Definitions and Data Description}

\paragraph{Laboratory Measurements (\( Y_i \))}
For each vegetable sample, we have laboratory measurements denoted as \( Y_i \), which include various physical metrics such as elastic resistance, weight change, luminosity, and size change, as described in \cite{Migal}.

\paragraph{Image Acquisition (\( Z_i \))}
The image dataset consists of RGB images of the vegetables, capturing essential visual attributes such as color, texture, and shape, with dimensions \( N \times M \). These images serve as the observable input data.

\paragraph{Quality Attribute (\( x_i \))}
The quality attribute of interest, \( x_i \), is the moisture content of the vegetable, represented as a two-dimensional matrix of size \( N \times M \), corresponding to the spatial distribution of moisture across the vegetable sample.

\paragraph{Moisture Content and Physical Laws}
The moisture content evolution is governed by Fick's second law of diffusion, which models the spatial and temporal changes in moisture concentration within the vegetable tissue. Specifically, the law is applied in a two-dimensional Cartesian coordinate system, as described by:
\[
\frac{\partial x}{\partial t} = D \left( \frac{\partial^2 x}{\partial u^2} + \frac{\partial^2 x}{\partial v^2} \right),
\]
where \( x(u, v, t) \) represents moisture concentration at spatial coordinates \( u \) and \( v \) over time \( t \), and \( D \) is the diffusion coefficient. This equation serves as the basis for the physics-guided inverse function \( \hat{g}^{-1} \), which translates image data into estimates of moisture content, ensuring that our model's predictions align with established physical principles.

\subsection{Methodological Adaptation for Thermal Camera Images}
\subsubsection{Overview}
The second case study applies our framework to a classification task involving images captured by a thermal camera. This dataset comprises 14,860 images of 15 different indoor materials and 26,584 images of 17 different outdoor materials. The physical law guiding our analysis is the heat equation, which models the distribution and flow of heat through materials. Thermal properties, such as conductivity, diffusivity, and specific heat capacity, vary significantly across different material types and influence their thermal behavior when exposed to heat sources. By leveraging the heat equation, we can relate the spatial temperature variations observed in the thermal images to the intrinsic properties of the materials, providing a physically-informed basis for enhancing classification accuracy.

The heat equation is suitable for this analysis because it describes how temperature evolves over time and space within a medium, making it applicable to both indoor and outdoor materials under various environmental conditions. For example, different materials exhibit distinct thermal patterns when subjected to heat, such as insulation properties or heat retention rates, which are captured by the thermal camera. These patterns are essential for distinguishing between material types.
The dataset utilized in this work was obtained with the agreement of the author from the original manuscript \cite{Deep_Thermal_Imaging}.
\subsubsection{Definitions and Data Description}

\paragraph{Target Labels (\( Y_i \))}
In this case, the target variable \( Y_i \) represents the material classification labels. These labels identify each image as belonging to one of the 15 indoor or 17 outdoor material classes.

\paragraph{Thermal Image Data (\( Z_i \))}
The image dataset \( Z_i \) consists of thermal images captured by an infrared camera, with dimensions \( N \times M \). These images reflect the temperature distribution across the surfaces of the materials, providing rich information on the thermal properties of each material type.

\paragraph{Latent Thermal State (\( x_i \))}
The latent variable \( x_i \) represents the underlying temperature distribution across the material's surface, modeled as a two-dimensional matrix of size \( N \times M \). This latent state captures how heat propagates through the material, reflecting its thermal properties and behavior.

\paragraph{Heat Transfer and Physical Laws}
The temperature distribution \( x \) evolves according to the heat equation, a fundamental physical law that describes heat transfer over time. In two-dimensional Cartesian coordinates, the heat equation is expressed as:
\[
\frac{\partial x}{\partial t} = \alpha \left( \frac{\partial^2 x}{\partial u^2} + \frac{\partial^2 x}{\partial v^2} \right),
\]
where \( x(u, v, t) \) represents the temperature at spatial coordinates \( u \) and \( v \) over time \( t \), and \( \alpha \) denotes the thermal diffusivity of the material. This equation captures how heat diffuses across the material’s surface, serving as a critical component for the inverse mapping function \( \hat{g}^{-1} \).

By incorporating the heat equation into our model, \( \hat{g}^{-1} \) is designed to translate the thermal image data \( Z_i \) into estimates of the latent temperature distribution \( x \). This physics-guided approach ensures that the classification model leverages the fundamental thermal properties of each material, resulting in improved recognition accuracy and a more interpretable understanding of the underlying heat transfer processes.

\subsection{Neural Network Architectures for Mapping Functions}

In our physics-guided framework, we utilize three distinct neural network models tailored to handle the complex relationships between image data and the target measurements. Each model is designed to align with the specific prediction strategies introduced in our methodology, ensuring that the neural networks are both effective and consistent with the underlying physical principles of the problem.

\paragraph{Model \( \hat{h} \) and \( \hat{f} \) - ResNet-18 Based Architectures}
The models \( \hat{h} \) (for direct prediction) and \( \hat{f} \) (for the second stage of inverse prediction) are both built upon the ResNet-18 architecture, a member of the Residual Network (ResNet) family of deep convolutional neural networks. ResNet-18 is renowned for its capability to tackle deep learning tasks efficiently, owing to the introduction of residual connections that mitigate the vanishing gradient problem typically encountered in deep networks. These skip connections facilitate the flow of gradients during backpropagation, enabling the network to learn more effectively, even with increased depth.

ResNet-18 consists of 18 layers with a series of convolutional blocks, each followed by activation and pooling layers. This architecture is particularly advantageous for our application, as it offers a balance between model complexity and computational efficiency, making it suitable for capturing the intricate features present in image data.

To adapt ResNet-18 for our regression tasks, where the aim is to map high-dimensional image data to scalar outputs \( Y_i \), we modified the architecture by introducing a tailored fusion block. By default, ResNet-18 produces a 512-dimensional feature vector in its final layer, which is typically suited for classification tasks. However, for regression purposes, this output requires further processing. Hence, we appended two fully connected layers to the architecture, forming a fusion block that transforms the 512-dimensional feature vector into the desired output space.

The enhanced architecture serves two distinct roles:
- For model \( \hat{h} \), this modification enables the direct mapping from image data \( Z_i \) to the laboratory measurements \( Y_i \).
- For model \( \hat{f} \), the fusion block facilitates the mapping from the intermediate quality attribute \( \hat{x} \), estimated from the image data via \( \hat{g}^{-1} \), to the final laboratory measurements \( Y_i \).

By integrating this fusion block, we effectively harness the feature extraction capabilities of ResNet-18 while adapting it for precise regression outcomes, ensuring that the models \( \hat{h} \) and \( \hat{f} \) are fine-tuned to handle the specific nature of our prediction tasks.

\paragraph{Model \( \hat{g}^{-1} \) - Physics-Guided CNN Transformation Network}
The Physics-Guided Mapping Function \( \hat{g}^{-1} \), responsible for the first stage of inverse prediction, employs a custom Convolutional Neural Network (CNN) Transformation Network. This model is designed to learn the mapping from the three-channel image data \( Z_i \) (e.g., RGB or thermal images) to a corresponding three-channel output representing the estimated quality attribute \( \hat{x}_i \). The architecture of \( \hat{g}^{-1} \) is carefully constructed to capture the spatial patterns and physical relationships embedded in the data, ensuring that the transformation adheres to the relevant physical laws.

The architecture of the CNN Transformation Network consists of the following key components:

\begin{itemize}
    \item \textbf{Convolutional Layers:} The network features a series of convolutional layers that sequentially extract spatial features from the input data. Each convolutional layer is followed by LeakyReLU activation functions to introduce non-linearity, and Batch Normalization to stabilize and accelerate the training process. The specific layers used are:
        \begin{itemize}
            \item \( \text{Conv2d}_{16} \): A convolutional layer with 16 output channels.
            \item \( \text{Conv2d}_{32} \): A convolutional layer with 32 output channels.
            \item \( \text{Conv2d}_{16} \): A second convolutional layer returning to 16 output channels.
        \end{itemize}
    These convolutional layers progressively capture complex spatial features while maintaining the input image's dimensionality, crucial for the task of reconstructing the output representation.
\begin{figure}[!ht]
\centering
\includegraphics[width=1\textwidth]{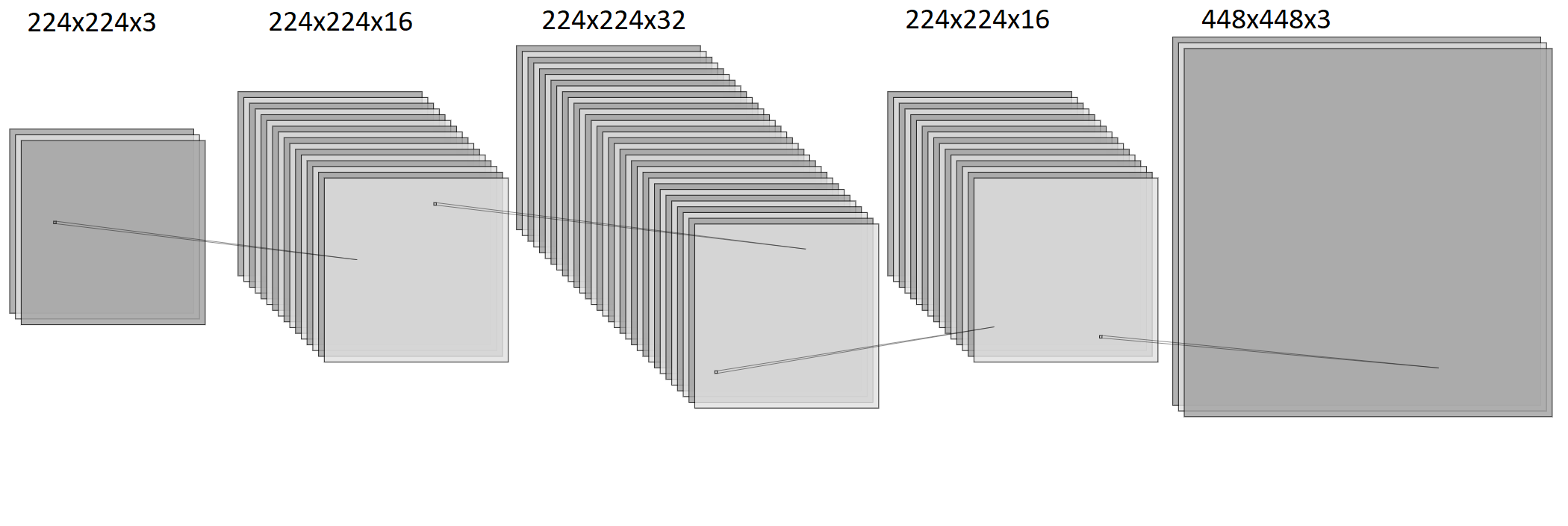}
\caption{The Physics-Guided Mapping Function \( \hat{g}^{-1} \) Architecture.}
    \label{fig:cnn}
\end{figure}
    \item \textbf{Dropout Regularization:} To prevent overfitting and improve the generalization capability of the model, a dropout layer with a rate of 0.2 is applied after each convolutional layer. This mechanism randomly drops a fraction of the neurons during training, ensuring that the network does not rely too heavily on specific features.

    \item \textbf{Transposed Convolution Layer:} The final layer of the network is a transposed convolution layer, \( \text{ConvTranspose2d}_3 \), which upscales the feature maps back to the original image size. This layer reconstructs the three-channel output, ensuring that the final estimated quality attribute \( \hat{x}_i \) retains the spatial resolution and structure of the input data.
\end{itemize}

The CNN Transformation Network \( \hat{g}^{-1} \) is specifically engineered to translate the input image data into an estimated quality attribute \( \hat{x}_i \) while maintaining adherence to the underlying physical laws (e.g., Fick's second law of diffusion or the heat equation). This design ensures that the network captures not only the visual features present in the input data but also the physical processes governing the system. The architecture effectively bridges the gap between empirical observations and theoretical modeling, making \( \hat{g}^{-1} \) a crucial component of the physics-informed prediction framework.

Figure \ref{fig:cnn} provides a detailed illustration of the CNN Transformation Network architecture, highlighting how each layer contributes to the overall transformation process.

\paragraph{Summary and Justification}
The combination of ResNet-18 based architectures for \( \hat{h} \) and \( \hat{f} \), along with the custom CNN Transformation Network for \( \hat{g}^{-1} \), allows us to construct a robust and flexible modeling framework capable of handling both direct and inverse prediction tasks. The use of residual connections in ResNet-18 facilitates deep feature extraction, while the tailored fusion block ensures that the output is appropriately adjusted for regression tasks. Meanwhile, the physics-guided CNN transformation effectively integrates domain knowledge into the learning process, resulting in a model that is not only data-driven but also informed by fundamental physical laws.

This carefully structured approach enables our neural network models to capture complex, non-linear relationships inherent in the data, leading to more accurate predictions and a deeper understanding of the underlying physical processes.

\subsubsection*{Definition of Adaptive Physics-Guided Neural Network (APGNN)}

The Adaptive Physics-Guided Neural Network (APGNN) integrates physical knowledge into neural network architectures by combining data-driven learning with domain-specific constraints. The APGNN framework consists of three neural components, \( \hat{h} \), \( \hat{g} \), and \( \hat{f} \), each performing distinct prediction tasks. Additionally, an adaptive weight function, \( \omega_i \), modulates the contributions of physics-based and data-driven predictions based on the physical alignment of the intermediate features.

\paragraph{\( \hat{h} \): Direct Prediction Function}  
\[
\hat{h}(Z_i; \theta_h) \equiv \text{NN}_h(Z_i; \theta_h)
\]
The direct prediction function \( \hat{h} \) maps the input features \( Z_i \) to the target prediction in a purely data-driven manner. While it captures complex non-linear patterns, it may not inherently respect underlying physical principles, risking overfitting.

\paragraph{\( \hat{g} \): Inverse Prediction Function}  
\[
\hat{g}(Z_i; \theta_x) \equiv \text{NN}_x(Z_i; \theta_x)
\]
The inverse prediction function \( \hat{g} \) predicts latent physical properties from the input data. This function leverages domain-specific knowledge, enhancing interpretability by reconstructing meaningful physical variables.

\paragraph{\( \hat{f} \): Forward Prediction Function}  
\[
\hat{f}(Z_i; \theta_y) = \text{NN}_y\left(\text{NN}_x(Z_i; \hat{\theta}_x); \theta_y\right)
\]
The forward prediction network \( \hat{f} \) maps the latent physical properties predicted by \( \text{NN}_x \) to the final target variables.

\paragraph{Adaptive Weight Function \( \omega_i \)}

The adaptive weight function \( \omega_i \) enables a dynamic balance between the physics-based and data-driven predictions, optimizing the final prediction based on the alignment of the intermediate representation \( \hat{x}_i \) with physical principles. This approach allows the model to adjust its reliance on the two sources of information according to the statistical relevance of the physics-based prediction, thereby leveraging both structured (physics-driven) and flexible (data-driven) components.

Specifically, \( \omega_i \) is computed as:
\[
\omega_i = T_{\omega}\left( \mathcal{L}_{x,n}(\text{NN}_x(Z_i; \hat{\theta}_x)) \right),
\]
where \( \mathcal{L}_{x,n}(\text{NN}_x(Z_i; \hat{\theta}_x)) \) quantifies the physical consistency of the intermediate state \( \hat{x}_i \) and acts as a proxy for model fit to the physical law. This design can be interpreted statistically as a form of data-driven regularization: when \( \hat{x}_i \) aligns well with physical laws, \( \omega_i \) favors the physics-based model, reducing variance by reinforcing domain consistency. Conversely, when alignment is weak, \( \omega_i \) shifts weight toward the data-driven model, allowing greater flexibility and reducing bias where the physics model may be less reliable.

The function \( T_{\omega} \) applies min-max normalization followed by a sigmoid transformation to map the result to the interval \( (0, 1) \):
\[
T_{\omega}(v) = \sigma\left( \frac{v - \min(v)}{\max(v) - \min(v) + \epsilon} \right), \quad \sigma(u) = \frac{1}{1 + e^{-u}},
\]
where \( \epsilon \) is a small positive constant to avoid division by zero. This transformation ensures that \( \omega_i \) lies within the range \( (0, 1) \), offering a probabilistic interpretation: \( \omega_i \) reflects the model’s confidence in the physical consistency of \( \hat{x}_i \), dynamically shifting reliance on either prediction pathway based on the observed data.

\paragraph{Final Prediction with Adaptive Balancing}
The final output \( \hat{y}_i \) is a weighted combination of the data-driven direct prediction \( \hat{h}(Z_i) \) and the physics-guided prediction \( \hat{f}(\hat{g}^{-1}(Z_i)) \):
\[
\hat{y}_i = \omega_i \hat{h}(Z_i) + (1 - \omega_i) \hat{f}(\hat{g}^{-1}(Z_i)).
\]
This weighted combination can be interpreted as a statistical ensemble model where each prediction source data-driven and physics-guided contributes based on its estimated reliability for the given input. By dynamically adapting \( \omega_i \) to data characteristics, the model balances bias and variance optimally, using the physics-guided approach when it is reliable and the data-driven model otherwise. This method mirrors the rationale of Bayesian model averaging, where the weighting adapts based on the fit quality, allowing the APGNN to flexibly adapt to diverse data scenarios and produce predictions that are both accurate and physically consistent.

\paragraph{Summary}
The APGNN architecture combines the strengths of data-driven neural networks with the interpretability and rigor of physics-based models. The inverse and forward networks \( \hat{g} \) and \( \hat{f} \) enforce physical consistency, while the direct neural network \( \hat{h} \) captures complex patterns in the data. The adaptive weight function \( \omega_i \) ensures that the final prediction remains robust, balancing the contributions of the two approaches according to the physical relevance of the intermediate representation.

\subsubsection*{Loss Function Incorporating Steady-State Assumption}
In line with the steady-state assumption, the neural network \( NN_x \), which serves as the Physics-Guided Mapping Function \( \hat{g}^{-1} \), is trained to estimate the quality attribute \( x \) (e.g., moisture content or temperature distribution) from the image data \( Z_i \). Under the steady-state assumption, both Fick's second law of diffusion and the heat equation reduce to the condition \( \Delta x = 0 \), implying that the Laplacian of \( x \) should be zero. Therefore, to ensure \( NN_x \) adheres to this physical constraint, the loss function \( \mathcal{L}_{x,n}(\theta_x) \) is formulated as:
\[
\mathcal{L}_{x,n}(\theta_x) = \frac{1}{n} \sum_{i=1}^n \left| \left( \frac{\partial^2 NN_x(Z_i; \theta_x)}{\partial u^2} + \frac{\partial^2 NN_x(Z_i; \theta_x)}{\partial v^2} \right) \right|^2,
\]

where \( \theta_x \) represents the parameters of the neural network \( NN_x \). This term ensures that the output of \( NN_x \) satisfies the steady-state condition by minimizing the deviation from \( \Delta x = 0 \) across all spatial coordinates \( u \) and \( v \).

To avoid trivial solutions where \( NN_x \) converges to an overly smoothed or constant output, an additional penalty term is introduced based on the inverse gradient magnitude. This term acts as a {variance-preserving mechanism}, ensuring that the model captures sufficient detail while conforming to the physical constraint. The total physics-informed loss function, \( \mathcal{L}_{physical}(\theta_x) \), thus becomes:

\[
\mathcal{L}_{physical}(\theta_x) = \left(\exp(\mathcal{L}_{x,n}(\theta_x)) - 1\right) + \frac{1}{n} \sum_{i=1}^n \left( \frac{1}{|\nabla_u NN_x(Z_i; \theta_x) + \epsilon| } + \frac{1}{|\nabla_v NN_x(Z_i; \theta_x) + \epsilon|} \right),
\]
where \( \nabla_u \) and \( \nabla_v \) denote the gradients in the \( u \) and \( v \) directions, respectively, and \( \epsilon \) is a small constant to avoid division by zero. The exponential transformation of \( \mathcal{L}_{x,n}(\theta_x) \) within the first term serves as an aggressive regularization, penalizing deviations from steady-state assumptions while retaining detail in regions where gradient changes are necessary.

By balancing bias and variance in this way, \( \mathcal{L}_{physical}(\theta_x) \) encourages \( NN_x \) to estimate the quality attribute \( x \) accurately while conforming to steady-state physical principles. This design choice adds statistical robustness by addressing the {bias-variance trade-off}, producing outputs that are both physically consistent and adaptable to inherent data variability.

\subsubsection*{Loss Function for Quality Metric Prediction}
In the Inverse Prediction approach, after estimating the quality attribute \( \hat{x} \) using \( NN_x \), a separate neural network \( NN_y \) (representing \( \hat{f} \)) is used to predict the laboratory measurements \( Y_i \) based on \( \hat{x} \). The loss function \( \mathcal{L}_{y,n}(\theta_y) \) for \( NN_y \) is structured as:
\[
\mathcal{L}_{y,n}(\theta_y) = \frac{1}{n} \sum_{i=1}^n \left| Y_i - NN_y(NN_x(Z_i;\hat{\theta}_x); \theta_y) \right|^2,
\]
where \( \theta_y \) denotes the parameters of \( NN_y \), and \( \hat{\theta}_x \) represents the optimized parameters of \( NN_x \). This loss function measures the deviation between the actual laboratory measurements \( Y_i \) and the predictions generated by \( NN_y \) based on the estimated quality attribute \( \hat{x}_i \).

Optimizing both \( \mathcal{L}_{physical}(\theta_x) \) and \( \mathcal{L}_{y,n}(\theta_y) \) ensures that the estimated quality attribute \( \hat{x}_i \) is not only consistent with the underlying physical laws but also leads to accurate predictions of the laboratory measurements \( Y_i \), thereby achieving a physics-constrained model that effectively integrates image data with empirical observations.


\section{Simulation}

To evaluate the effectiveness of our physics-guided learning framework, we conducted simulations on synthetic data generated using three distinct governing differential equations. These simulations emulate complex physical phenomena in 2D domains, providing a variety of spatial patterns for training and testing the neural network models. Additionally, multiple non-linear transformations were applied to these generated fields to create diverse image representations, mimicking real-world variations often encountered in physical systems.

For the simulation study, synthetic images were generated using governing equations, with a fixed seed for each set to ensure reproducibility. The datasets were split into training and testing sets, with training sizes of 45, 70, 85, 100, 200, and 300 images, enabling an evaluation of model performance across different sample sizes. A consistent static test set of 100 images was used throughout the study to provide a standardized benchmark for model accuracy. This entire process was repeated across 100 synthetic datasets, ensuring a robust assessment of the model’s learning ability and stability under various simulated conditions.
\subsection{Generating 2D Domains with Governing Equations}

The synthetic data was generated using the following three governing differential equations, which simulate different physical processes:

\paragraph{1. Transient 2D Diffusion Equation}
The first simulation scenario was based on the transient 2D diffusion equation, governed by:
\[
\frac{\partial x}{\partial t} = D \left( \frac{\partial^2 x}{\partial u^2} + \frac{\partial^2 x}{\partial v^2} \right),
\]
where \( D \) is the diffusion coefficient, and \( x(u, v, t) \) represents the evolving physical field over a 2D spatial domain. 

The initial condition \( X_p \) was generated using a bilinear interpolation of random boundary values, and the equation was evolved over time using an explicit finite difference method. Zero Dirichlet boundary conditions were applied, resulting in a dynamic diffusion process that smooths out the initial field over time. This simulation captures the transient behavior of diffusive systems and provides a realistic representation of physical phenomena such as heat distribution or moisture diffusion.

\paragraph{2. Advection-Diffusion Equation with Source Term}
In the second scenario, an advection-diffusion equation was used to generate the domain. A random initial condition \( u \) was evolved over multiple time steps, incorporating both diffusion and a spatially varying source term:
\[
\frac{\partial u}{\partial t} = D \nabla^2 u + \text{source},
\]
where \( \text{source} = \sin(X) \cos(Y) + 0.5 \eta \) with \( \eta \sim \mathcal{N}(0, 1) \). The solution was further refined using a Gaussian filter to smooth out artifacts, and periodic boundary conditions were applied to ensure continuity at the domain edges.

\paragraph{3. Poisson Equation with Complex Source Term}
The third case involved solving a Poisson equation with a non-linear, spatially varying source term. The source term was constructed by combining sinusoidal patterns and Gaussian peaks, adding significant complexity to the generated fields:
\[
\nabla^2 \phi = -\text{source term}.
\]
The solution was iteratively computed using a relaxation method, ensuring convergence. This produced intricate spatial distributions resembling potential fields.

For all generated domains, the final label \( y_i \) was calculated as the sum of the original physical properties \( x_i \) over the grid, with added Gaussian noise to simulate measurement uncertainties.

\subsection{Non-linear Transformation for Image Representation}

To simulate the transformation from physical properties to image data, we applied a non-linear mapping function to the generated 2D domains. This transformation emulates the complex processes by which real-world image data represent underlying physical phenomena.

In this context, \( s \) denotes the spatial coordinates within the 2D domain \( \Omega \). At each spatial location \( s \), the physical property of interest is represented by \( x(s) \), which could correspond to quantities such as temperature, concentration, or moisture content, depending on the governing physical process. The objective of the non-linear transformation is to map these physical properties \( x(s) \) to an image representation \( z(s) \), simulating how these physical phenomena might appear in a real-world imaging scenario.

The non-linear transformation applied is defined as follows:

\[
z(s) = T\left(0.5 \cdot \exp(x(s)) + 0.3 \cdot x(s)^2 + 0.2 \cdot \sqrt{x(s)}\right)
\]

This transformation uses a weighted combination of exponential, quadratic, and square root functions to create a complex and dynamic mapping from \( x(s) \) to \( z(s) \). This mapping captures diverse image features by highlighting high-value regions, preserving smooth transitions, and introducing subtle variations, thereby simulating the nuanced ways in which physical phenomena can manifest in real-world image data.

\subsubsection*{Intermediate Transformation \( T \)}
The transformation \( T \) is designed to convert the underlying physical property into a visually interpretable RGB image. The process involves several steps to ensure that the resulting image representation effectively highlights key features encoded in the transformed physical data.

First, the transformed data is normalized to the range \([0, 1]\) to ensure consistency across different inputs. This normalization is essential, as it rescales the data to remove the effect of varying magnitudes and facilitates proper mapping to color values.

Next, the normalized data is mapped to an RGB color space using a colormap. The default colormap used is the 'plasma' colormap from Matplotlib. However, a custom colormap can also be applied. The custom colormap is defined by linearly interpolating between two colors: dark green \([0, 0.5, 0]\) and yellow \([1, 1, 0]\). This interpolation creates a smooth gradient from yellow to dark green, ensuring a meaningful visual representation of the normalized data.

The RGB conversion step involves applying the selected colormap to the normalized data to generate an RGB image representation. The resulting RGB values are then scaled to an 8-bit color format (i.e., each color channel takes values between 0 and 255). Additionally, any alpha channel present (used for transparency) is removed to ensure the output is strictly in RGB format. 

This RGB image representation serves as a visual encoding of the underlying physical property, enabling the network to process the transformed image data effectively. By employing both standard and custom colormaps, the transformation ensures flexibility in how the physical properties are visualized, enhancing interpretability in different contexts.
\subsection{Label Calculation and Introduction of Noise}

For each synthetic data sample, we define two types of labels to support both regression and classification tasks. The continuous label \( y \) for regression is computed as the integral of the physical property \( x(s) \) over the spatial domain \( \Omega \):
\[
y = \int_{\Omega} x(s) \, ds,
\]
where \( s \) denotes the spatial coordinates within \( \Omega \). This integral captures the total accumulated value of \( x(s) \) across the domain, acting as the target for the regression task in our neural network's training.

For the classification task, the continuous \( y \) values were divided into 5 equal intervals, generating distinct classes. Each class corresponds to a quantile range, effectively grouping similar levels of \( y \) to create ordinal labels that represent varying intensities or accumulations of the physical property \( x(s) \). This setup provides a clear separation for classification while maintaining relevance to the underlying physical processes.

Gaussian noise was added to the original domains and the continuous labels \( y_i \), simulating real-world measurement variability. The noise was introduced post-calculation of \( y_i = \sum x_i \) but prior to the image transformations. This additional complexity helps to more accurately evaluate the robustness and predictive power of the neural network models under realistic conditions.

By generating synthetic 2D domains using multiple governing physical equations and applying non-linear mappings, we created a diverse dataset for rigorous evaluation. This dataset enables testing of our physics-guided learning approach across both regression and classification tasks, ensuring that the neural network \( \hat{g}^{-1} \) can map image data \( Z_i \) back to the underlying quality attribute \( x_i \) in a manner that respects the governing physical laws.

\subsection{Results and Discussion of the Simulation Study}

This section presents the simulation results, comparing the performance of the Adaptive Physics-Guided Neural Network (APGNN), Physics-Guided Neural Network (PGNN), and a standard data-driven ResNet approach across different governing equations and a single non-linear transformation. The models were assessed on both regression and classification tasks, with evaluation metrics including average Root Mean Square Error (RMSE) for regression and F1 score for classification, each plotted against various training set sizes.

Across regression tasks, the APGNN and PGNN consistently demonstrated better predictive performance compared to the standard ResNet, with lower RMSE values observed across all governing equations, showing that physics-based guidance enhances prediction accuracy for continuous target variables. Interestingly, in regression tasks, the influence of the underlying governing equation was minimal across the PGNN and APGNN, suggesting that both models adapt well to diverse physical processes by incorporating generalizable physical constraints into their architecture.

For classification tasks, however, the ResNet model exhibited competitive performance, especially under complex physical scenarios represented by the advection-diffusion and Poisson equations. The APGNN still achieved superior F1 scores, indicating that the adaptive weighting of physics-guided and data-driven predictions helps it manage non-linear boundaries effectively. In contrast, the PGNN, which solely relies on a physics-based constraint, did not perform as well as the APGNN or ResNet in classification. This result underscores a key difference in the role of physics guidance between regression and classification tasks: while regression tasks benefit directly from the physics-based smoothness and continuity constraints, classification tasks often require handling intricate decision boundaries, for which a purely data-driven model like ResNet is more naturally suited.

In summary, the APGNN consistently outperforms both PGNN and ResNet across tasks, effectively balancing physics-based and data-driven elements for both regression and classification. These findings illustrate that while PGNN provides advantages in scenarios where physical laws are critical, the APGNN’s adaptive framework offers enhanced flexibility and robustness across varying tasks and physical conditions.

\begin{figure}[ht]
    \centering
    \includegraphics[width=\textwidth]{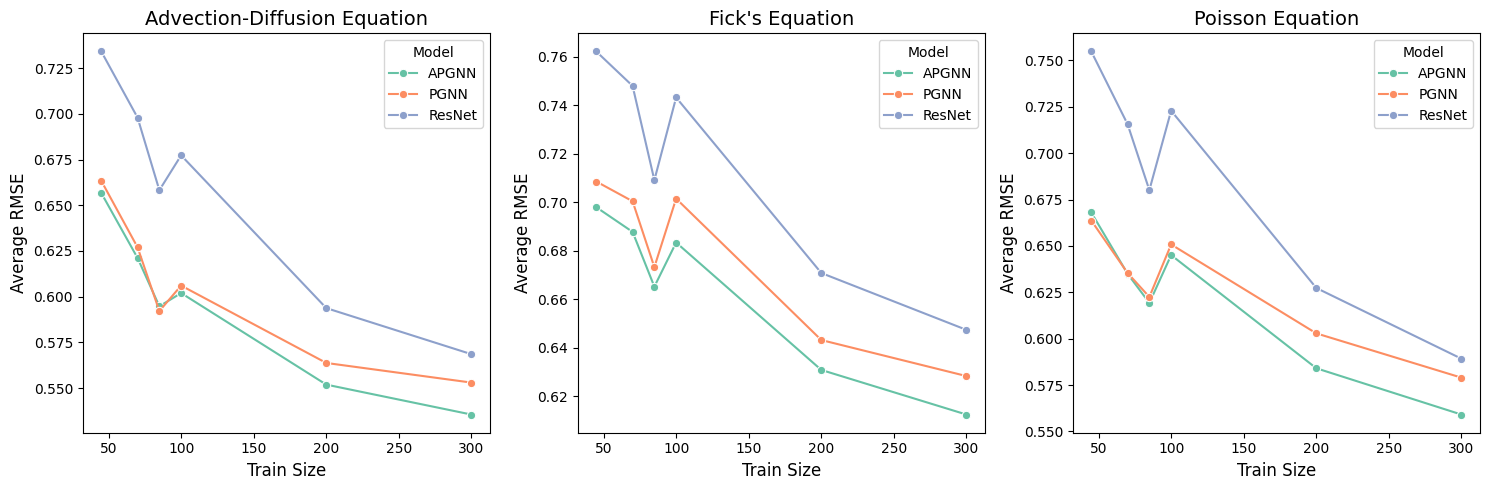}
    \caption{Average RMSE Score vs. Training Size for regressions task.}
    \label{fig:simulation0}
\end{figure}

\begin{figure}[ht]
    \centering
    \includegraphics[width=\textwidth]{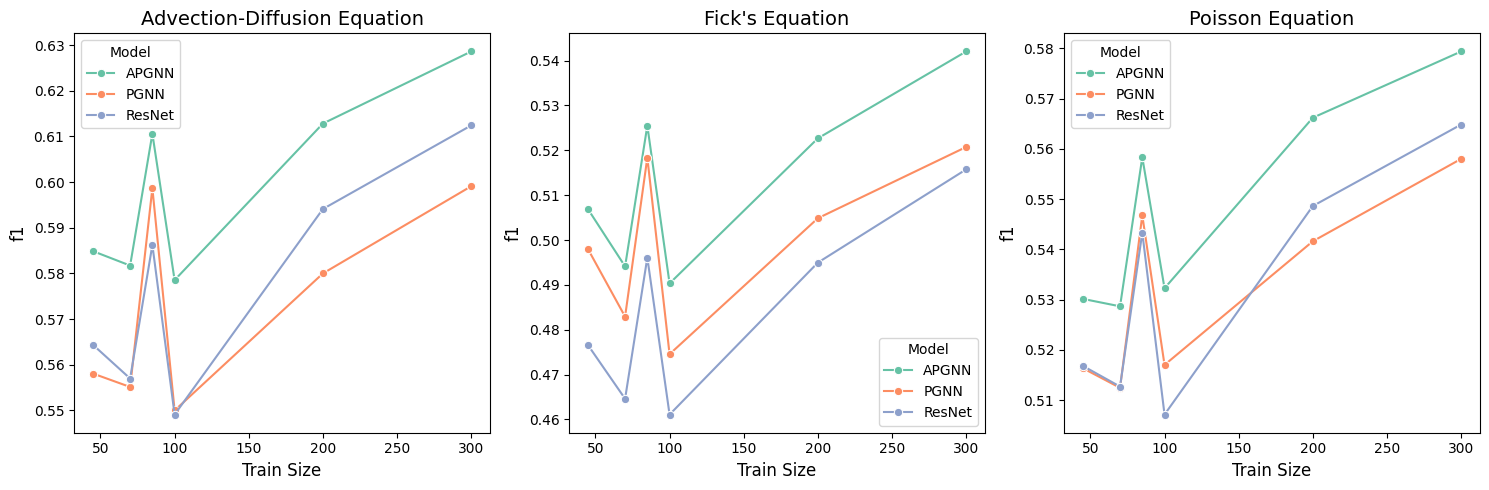}
    \caption{Average F1 Score vs. Training Size for classifications task.}
    \label{fig:simulation3}
\end{figure}

\paragraph{Visual Representation of Results}
Figures \ref{fig:simulation0},  illustrate the comparative performance of the PGNN, ResNet, and APGNN approaches for the regression task, with RMSE values plotted across varying training set sizes. Figure \ref{fig:simulation3} provides a similar comparison for the classification task, displaying F1 scores across training sizes. Each set of figures includes subplots corresponding to each governing equation and non-linear mapping function, enabling a comprehensive analysis of model performance for both regression and classification tasks.

\paragraph{Summary of Findings} 

The simulation study reveals distinct advantages of incorporating physics-guided constraints in predictive modeling, with the APGNN consistently achieving superior performance across both regression and classification tasks. For regression tasks, both the APGNN and PGNN outperform the data-driven ResNet model, with minimal performance differences observed across different governing equations. This suggests that incorporating physical constraints into neural networks can improve prediction accuracy in continuous-valued tasks, likely due to the smoother, physically consistent representations enforced by the Laplacian-based loss function.

In contrast, for classification tasks, the ResNet model showed competitive performance, particularly for complex governing equations such as the advection-diffusion and Poisson equations. While the APGNN continued to outperform other models, the purely physics-guided PGNN struggled in this context, emphasizing that the value of physics-based constraints may diminish when modeling intricate decision boundaries, as required in classification. The adaptive weighting in APGNN allows it to balance physics-based and data-driven predictions, making it better suited to a variety of tasks and physical scenarios.

These findings highlight the strengths of adaptive, physics-informed frameworks like the APGNN, which leverage physical knowledge while remaining flexible across different problem types. In summary, physics-guided learning shows clear benefits in regression scenarios with continuous outputs, while an adaptive approach offers robust performance across both regression and classification tasks.



\section{Real Data}

To evaluate the performance of our adaptive physics-guided neural network (APGNN) approach on real-world scenarios, we conducted experiments using two different datasets: a cucumber image dataset with continuous fitness labels and a thermal camera material classification dataset. These datasets allowed us to assess the efficacy of our model in both regression and classification contexts.

\subsection{Cucumber Image Dataset}

The cucumber dataset consists of 660 RGB images captured post-storage, exhibiting natural variations in color (ranging from green to yellow) and texture. Each image was preprocessed to focus on a standardized Region of Interest (ROI) measuring \(110 \times 350\) pixels. To enhance the image quality and reduce noise, Gaussian blur was applied, and data augmentation techniques such as resizing, random vertical and horizontal flips, and normalization were employed. These steps ensured a more diverse and robust dataset for training.

The primary objective was to predict cucumber quality, which was labeled on a continuous scale from 0 to 17, based on laboratory measurements of elastic resistance, luminosity, weight change, and size, as described in \cite{Migal}. This label serves as the ground truth for a regression task, and the process of inverse regression was employed to emphasize moisture-relevant attributes in the images, ensuring that they align with the APGNN's focus on analyzing moisture content.

For model training, the neural networks \( \hat{h} \), \( \hat{g}^{-1} \), and \( \hat{f} \) were optimized using a learning rate of 0.0001. The Physics-Guided Mapping Function \( \hat{g}^{-1} \) was trained for 25 epochs, after which it was set to evaluation mode. The direct prediction model \( \hat{h} \) and the inverse prediction model \( \hat{f} \) were then trained for 55 epochs, with the final performance evaluated as the mean across epochs 25 to 55. Stratified splitting ensured balanced label distribution across training and test sets.

To comprehensively evaluate model performance, we used a static test set of 200 samples as a benchmark for predicting cucumber quality. In addition, training sets of varying sizes (50, 80, 100, 150, 200, 250, 350, and 400 samples) were employed to examine how the model's accuracy scales with the amount of training data available. This approach allowed for an in-depth analysis of the model's learning efficiency.

We compared three approaches: Direct Prediction (\( \hat{h} \)), Inverse Prediction (\( \hat{g}^{-1} \) and \( \hat{f} \)), and the Adaptive Prediction strategy. The hypothesis posited that the Inverse Prediction model would outperform the Direct Prediction approach, particularly when trained on smaller datasets. The performance metrics used for evaluation included RMSE, MAE, standard errors, and \( R^2 \) values, providing a comprehensive assessment of each model's predictive capability for cucumber quality.

\subsection{Thermal Camera Material Classification Dataset}

The second dataset involves thermal camera images, designed for material classification tasks, and is divided into two distinct subsets: indoor and outdoor materials. This dataset was introduced as part of the "Deep Thermal Imaging" study \cite{Deep_Thermal_Imaging}.

The indoor subset contains 14,860 images representing 15 different material types found in indoor environments, such as paper, MDF, nylon stiff fabric, and bubble wrap. These materials were chosen based on their common presence in households or office settings. The outdoor subset consists of 26,584 images capturing 17 distinct outdoor materials, such as brick walls, asphalt roads, granite kerb stones, and flower beds. These materials represent the varied textures and surfaces commonly found in outdoor environments, and their thermal patterns can be influenced by weathering, dirt, and exposure to environmental conditions.

For the Thermal Camera Material Classification Dataset, a static test set consisting of 4,000 images was utilized, providing a consistent benchmark for evaluating model performance. The training sets were prepared with the following sizes: 500, 1,000, 2,000, 3,500, 5,000, 6,500, 7,000, 8,000, and 10,000 images. To ensure reproducibility, the split was performed using fixed seed values. This process was repeated across 100 different iterations of the dataset, allowing for a comprehensive analysis of the model's performance and consistency under varying training sizes.

Thermal images were captured using a low-cost, mobile thermal camera (FLIR One 2G), which was integrated into a smartphone to capture the spatial temperature patterns of each material. The dataset collection was conducted under different illumination settings (daytime/nighttime), varying environmental temperatures, multiple viewpoints, and at proximate distances ranging from 10 cm to 50 cm from the camera lens. This ensured that the dataset captured a wide variety of thermal textures, enhancing the robustness of the material recognition models.

The thermal images were preprocessed using a dynamic range quantization technique to amplify the temperature differences across the material surfaces, making the textures more discernible and invariant to changes in ambient temperature \cite{Deep_Thermal_Imaging}.

The dataset serves as a comprehensive benchmark for testing the effectiveness of the APGNN approach in classifying materials based on their thermal properties. It provides a challenging testbed for evaluating the model's ability to generalize across different material types and environmental conditions.

\subsection{Results}

\begin{figure}[ht]
    \centering
    \includegraphics[width=0.7\textwidth]{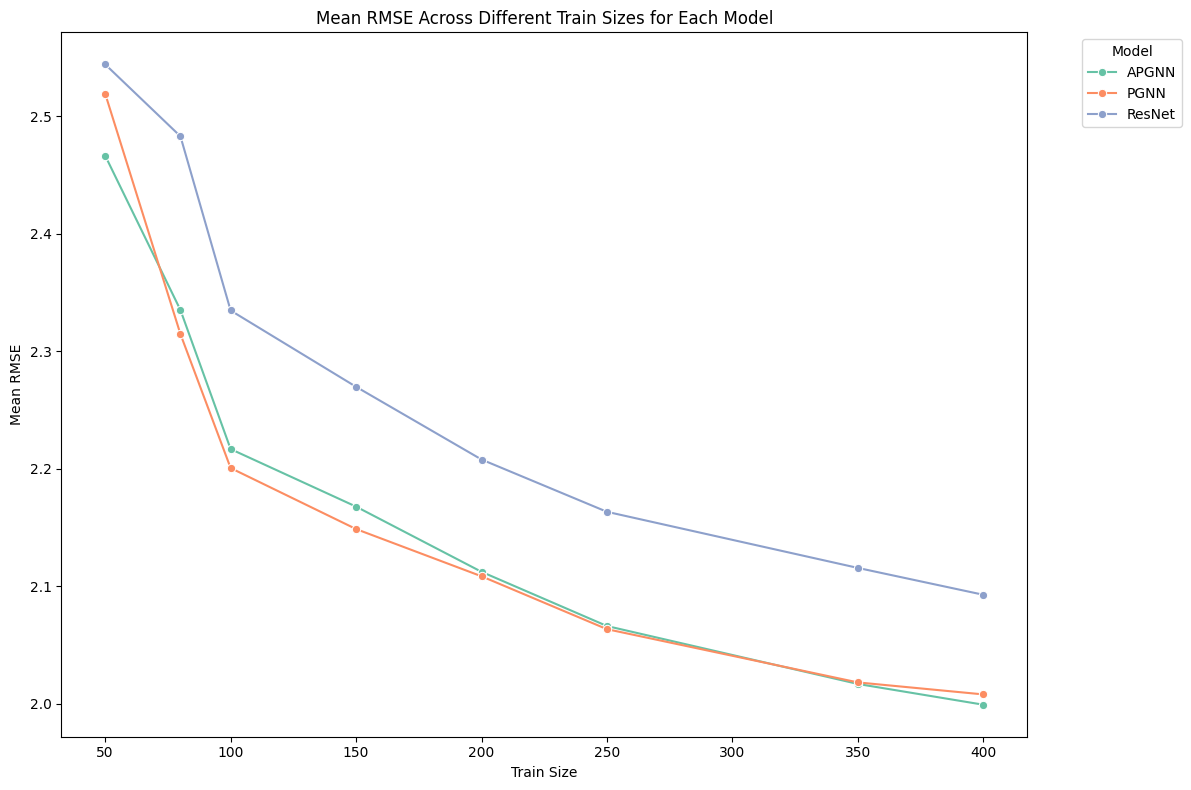}
    \caption{Mean RMSE across different training sizes for each model on the cucumber dataset. The PGNN consistently outperforms the ResNet model, with the Adaptive Prediction approach showing the best performance overall.}
    \label{fig:cucumber_rmse}
\end{figure}

\paragraph{Cucumber Dataset}
Figure \ref{fig:cucumber_rmse} shows the Mean RMSE across different training sizes for each model on the cucumber dataset. The APGNN and PGNN approach consistently demonstrates better performance over the vanilla ResNet model across all training sizes, as indicated by the lower RMSE values. This result confirms the hypothesis that the physics-guided neural network can more effectively capture the underlying moisture-related attributes of the cucumber images, resulting in improved prediction accuracy.

Across all training sizes, the physics-guided approach achieves the lowest RMSE. However, the adaptive mechanism in APGNN does not significantly improve performance over the purely physics-guided approach (PGNN) in the cucumber dataset scenario. This suggests that in controlled environments with consistent physical properties, the purely physics-guided approach (PGNN) provides optimal performance, while the adaptive mechanism offers limited additional benefit.

\begin{figure}[ht]
    \centering
    \begin{minipage}{0.49\textwidth}
        \centering
        \includegraphics[width=\textwidth]{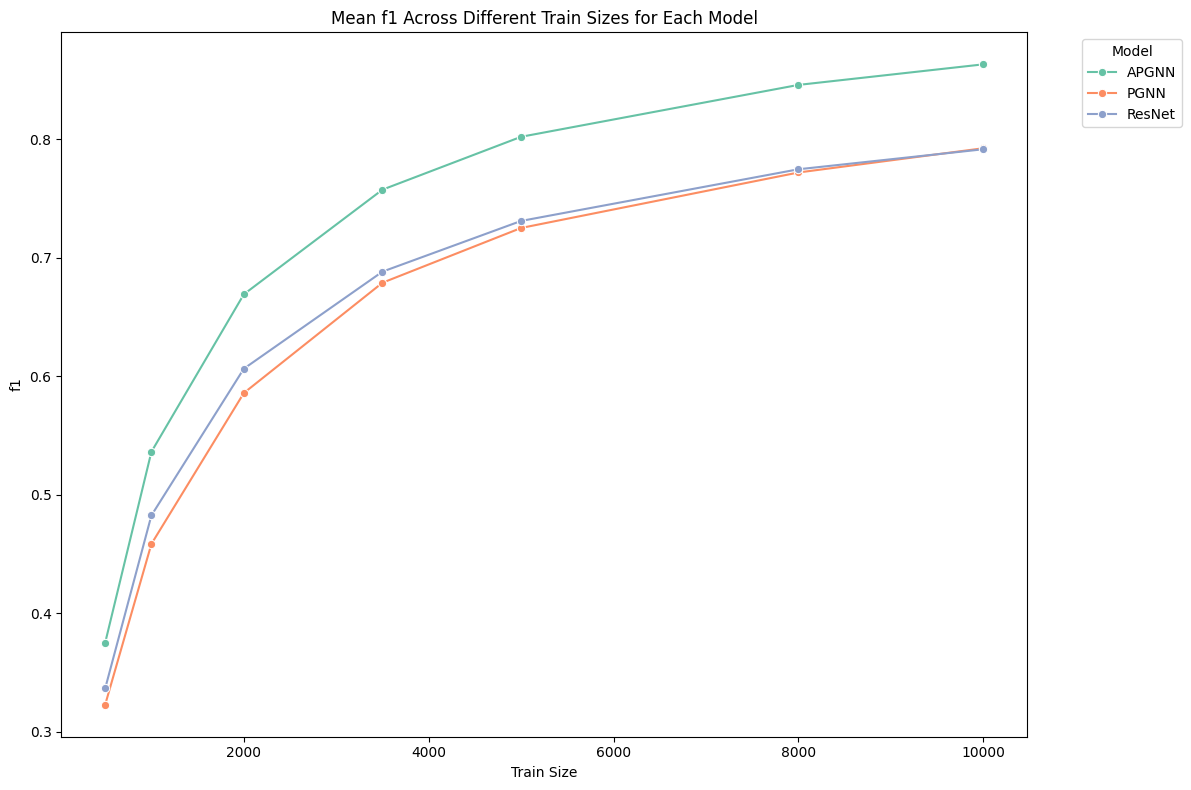} 
        \caption*{(a) Indoor Material Classification} 
    \end{minipage}
    \hfill
    \begin{minipage}{0.49\textwidth}
        \centering
        \includegraphics[width=\textwidth]{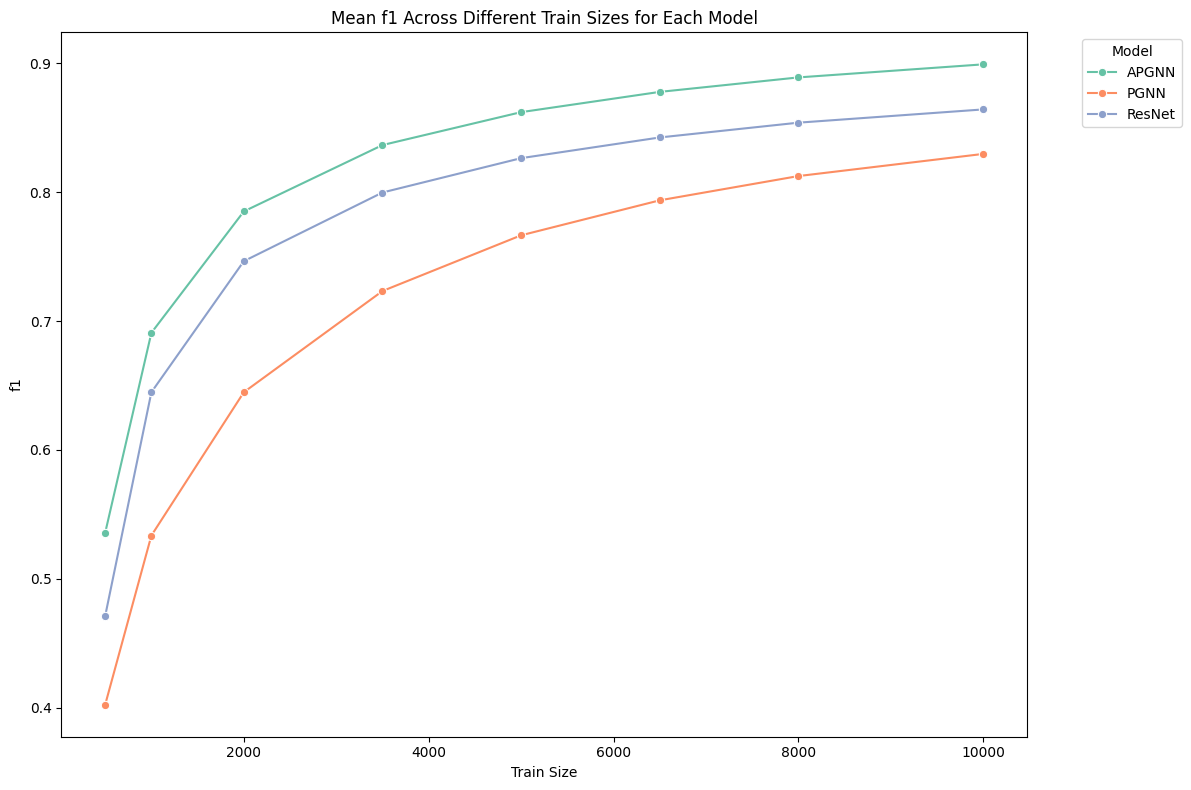}
        \caption*{(b) Outdoor Material Classification} 
    \end{minipage}
    \caption{Mean F1 across different training sizes for the indoor (left plot) and outdoor (right plot) material classification tasks using the Thermal Camera Material Classification Dataset. The ResNet model demonstrates better performance than the PGNN over all train sizes for outdoor images, and for indoor images until a training size of 8000 images, after which the PGNN achieves the same performance.}
    \label{fig:thermal_f1}
\end{figure}

\paragraph{Thermal Camera Material Classification Dataset}
The performance results for the Thermal Camera Material Classification Dataset are displayed in Figure \ref{fig:thermal_f1}, showing the Mean F1 scores across different training sizes for both indoor and outdoor material subsets. The side-by-side plots facilitate a direct comparison of model performance across these distinct environments.

In contrast to the cucumber dataset, where controlled conditions favored the physics-guided approach, the ResNet model consistently outperforms PGNN in outdoor material classification, demonstrating superior adaptability to the increased variability and noise in outdoor thermal images. For indoor materials, however, performance trends show a shift with increasing training size. Initially, ResNet achieves higher F1 scores, but as training size reaches approximately 8,000 images, the PGNN performance converges with ResNet, suggesting that a larger dataset allows PGNN to capture the underlying physical patterns more effectively.

Importantly, the adaptive mechanism in APGNN provides a significant improvement over both PGNN and ResNet across all training sizes. By dynamically weighting the contributions of the physics-based and data-driven predictions, APGNN adapts effectively to the different levels of noise and complexity in both indoor and outdoor datasets, resulting in the consistently highest performance. This demonstrates the advantage of combining physical insights with data-driven flexibility, particularly in varied and complex environments.

\subsection{Discussion}

The superior performance of the Physics-Guided Neural Network (PGNN) on the cucumber image dataset across all training sizes, as compared to the thermal camera images, can be attributed to several key factors, including the nature of the dataset, the diversity of material properties, and the controlled environment in which the images were captured.

Firstly, the cucumber dataset exhibited less diversity in material properties, as all the images represented a single type of produce (cucumbers) with relatively uniform characteristics. In contrast, the thermal camera dataset contained images of 15 different indoor materials and 17 different outdoor materials, each with unique thermal properties. This increased diversity introduces greater complexity and variability in the thermal images, making it more challenging for the PGNN to accurately model the underlying physical patterns. The homogeneous nature of the cucumber dataset allowed the physics-guided approach to leverage its assumptions about moisture diffusion, resulting in more accurate predictions and better alignment with the physics-guided loss.

Secondly, the cucumber images were captured under laboratory-controlled conditions, ensuring consistent lighting, temperature, and humidity. This stable environment minimizes the presence of noise and extraneous variables, allowing the PGNN to focus more effectively on relevant moisture-related features without interference from external factors. In contrast, the thermal images were collected in less controlled settings, particularly for outdoor materials, where factors such as changing weather, lighting conditions, wind, and ambient temperature introduce additional noise. These uncontrolled variations make it more difficult for the physics-guided PGNN to maintain an advantage, whereas the data-driven ResNet model shows better adaptability to these complex, noisy conditions.

The Adaptive Physics-Guided Neural Network (APGNN), which incorporates a dynamic weighting mechanism, was able to further improve prediction accuracy, particularly in the thermal camera dataset. By adaptively adjusting the reliance on physics-based or data-driven predictions, APGNN could capitalize on the physics-guided approach when physical consistency was high, while leveraging the flexibility of ResNet when conditions were less predictable. This adaptability was especially valuable for the diverse indoor and outdoor material classification tasks in the thermal dataset, where APGNN consistently outperformed both the PGNN and ResNet across training sizes. However, in the cucumber dataset, where conditions and properties were more stable and aligned with the physical model, the adaptive mechanism in APGNN did not yield significant improvements over the purely physics-guided PGNN.

In summary, the PGNN’s strong performance on the cucumber dataset can be attributed to reduced material diversity, a controlled acquisition environment, and the richer feature set provided by RGB imaging, which collectively create a favorable scenario for the physics-guided approach. In contrast, the APGNN’s dynamic weighting mechanism proved especially beneficial for the more complex thermal material classification tasks, enabling it to balance physics-based and data-driven predictions in response to environmental variability and material diversity. This study highlights the strengths of physics-guided learning in structured settings and the potential of adaptive models like APGNN to handle a wider range of real-world scenarios.

\paragraph{Environmental Stability and Noise Levels}
One of the primary reasons for the superior performance of the PGNN on indoor materials is the relatively stable and controlled environment in which these images were captured. Indoor settings are less prone to fluctuations in temperature, lighting, and other external variables. This stability ensures that the thermal patterns captured by the camera are more representative of the true material properties, allowing the PGNN to effectively leverage its physics-based constraints to model heat transfer and temperature distribution accurately.

In contrast, outdoor environments are inherently more variable and subject to a wide range of environmental factors such as changes in sunlight, wind, rain, and humidity. These fluctuations introduce significant noise and complexity into the thermal images, making it challenging for the PGNN to adhere strictly to the assumptions of the heat equation or steady-state diffusion model. As a result, the ResNet model, which is purely data-driven and more adaptable to varying patterns and noise, performs better in this outdoor context.

\paragraph{Influence of Multiple Physical Phenomena}
The PGNN's performance on indoor materials can be attributed to the fact that these materials are primarily governed by steady-state heat transfer processes, such as conduction, which align well with the underlying physical assumptions incorporated into the PGNN's loss function. This alignment enables the PGNN to accurately capture the thermal distribution patterns of indoor materials, leading to improved performance.

However, outdoor materials are influenced by a broader range of physical phenomena, including convection (e.g., wind effects), radiation (e.g., sunlight exposure), and even moisture absorption. These additional factors result in complex thermal behaviors that are not well captured by the steady-state heat equation or the Laplacian minimization approach employed by the PGNN. The ResNet model, with its purely data-driven architecture, is more flexible and capable of learning these complex, non-linear relationships from the data, which explains its superior performance in the outdoor setting.

\paragraph{Material Diversity and Complexity}
The indoor material dataset comprises a relatively homogeneous set of materials with well-defined and predictable thermal properties (e.g., paper, wood, leather). This uniformity allows the PGNN to leverage the physics-based constraints effectively, as the thermal behaviors are more consistent and aligned with the assumed physical model.

In contrast, the outdoor dataset includes a diverse array of materials, such as asphalt, brick, vegetation, and metal, each exhibiting distinct thermal properties. Moreover, these materials are subject to external influences such as weathering and environmental exposure, which further complicates the thermal patterns. The PGNN, which relies on the assumption of consistent heat transfer properties, struggles to generalize across such a diverse set of materials, while the ResNet model's data-driven approach enables it to capture the varying characteristics more accurately.

\paragraph{Implications and Future Work}

The findings from this study suggest that while the PGNN approach is highly effective in controlled, stable environments where physical constraints align well with the data (such as in indoor materials or uniform produce like cucumbers), it is less suited for scenarios where multiple physical phenomena and environmental variability introduce significant noise and complexity (such as outdoor settings). The superior performance of the ResNet model in outdoor environments underscores the importance of flexibility and adaptability when learning from diverse and noisy datasets.

These results indicate that the choice between physics-guided and purely data-driven approaches should be carefully informed by the environmental context and material properties, as each method offers unique advantages depending on the complexity and stability of the underlying physical patterns. The APGNN model demonstrates a promising avenue for combining the strengths of both approaches: by adaptively balancing physics-guided and data-driven components, APGNN achieves robust performance across a wide range of scenarios. This adaptability suggests potential for future applications where environmental conditions vary significantly or where physical laws partially capture, but do not entirely govern, the observed data.

Future work will focus on further refining the adaptive weighting mechanism in APGNN to enhance its responsiveness to changing physical conditions and environmental noise. Additionally, expanding APGNN to integrate other domain-specific physical laws could extend its applicability across diverse fields, potentially providing a unified framework for tasks where both physical consistency and adaptability are critical for accurate predictions.

\section{Conclusion}

In this study, we introduced the Adaptive Physics-Guided Neural Network (APGNN) framework for predicting quality attributes from image data by integrating fundamental physical laws directly into the deep learning process. The APGNN adaptively combines physics-informed and data-driven predictions through a dynamic weighting mechanism, allowing the model to balance its reliance on physical laws and neural predictions based on the data’s consistency with known physics. We evaluated this approach on both synthetic and real-world datasets, with comparisons to the Adaptive Physics-Guided Neural Network (APGNN), Physics-Guided Neural Network (PGNN) and a standard data-driven ResNet model.

The results of the simulation study demonstrated that APGNN consistently outperformed both PGNN and ResNet across all tasks and governing equations. By dynamically adjusting the balance between physics-based and data-driven predictions, APGNN was able to generalize effectively even across varying physical conditions.

For regression tasks in the simulation study, both APGNN and PGNN showed comparable performance, with outperforming ResNet across all governing equations. This advantage can be attributed to the physics-guided structure in PGNN, which aligns well with continuous output targets and helps capture the underlying physical laws governing the simulated systems. 

In contrast, for classification tasks, the Adaptive Physics-Guided Neural Network (APGNN) consistently outperformed both PGNN and ResNet approaches. However, ResNet often performed comparably to, or even surpassed, PGNN in settings where the physical dynamics diverged from the objectives of the physics-guided loss function, which focused on Laplacian minimization. While PGNN excelled in scenarios governed by the diffusion equation, where the physics-guided loss aligned directly with minimizing the Laplacian, this rigid adherence to physics-based constraints limited its flexibility in classification contexts. Classification tasks demand distinct categorization boundaries, which can conflict with PGNN's smooth, continuous approach rooted in physical principles. In contrast, ResNet’s purely data-driven structure allowed it to flexibly adapt to the complex, non-linear boundaries required for effective classification, particularly in settings governed by equations with non-zero Laplacian terms. This distinction highlights APGNN’s advantage in dynamically balancing both approaches, enabling it to maintain robust classification performance across various governing physical conditions.

In real-world experiments, APGNN consistently achieved the best results across datasets with varying complexity. On the cucumber dataset, characterized by low material diversity and stable conditions, APGNN demonstrated similar performance to PGNN, reflecting the effectiveness of physics-guided learning in controlled environments. In the more complex thermal camera dataset, with diverse indoor and outdoor materials, APGNN outperformed both PGNN and ResNet. Here, ResNet surpassed PGNN due to its purely data-driven approach, which allowed it to better handle the higher noise and variability present. APGNN, however, effectively balanced physics-based and data-driven insights, dynamically adjusting its reliance on each according to the conditions, leading to robust and superior performance across diverse environments.

These findings highlight the advantage of adaptive physics-guided frameworks like APGNN, which leverage both domain-specific physics and data-driven flexibility to handle a broad spectrum of regression and classification tasks. The APGNN framework’s dynamic weighting mechanism proves especially valuable for real-world applications where physical consistency is desired but data complexity and variability must also be accommodated.


\subsection*{Conflict of Interest}

The author has no conflicts to disclose.

\subsection*{Data Availability Statements}

The datasets generated and analyzed during this study can be obtained from the corresponding author upon reasonable request.

\bibliographystyle{jasa3}

\bibliography{references}
\end{document}